\documentclass[aps,prb,twocolumn,superscriptaddress,amsmath,amssymb]{revtex4-2}

\usepackage{graphicx}
\usepackage{dcolumn}
\usepackage{bm}

\begin{document}

\title{Resonance fluorescence spectral dynamics of an acoustically modulated quantum dot}

\author{Daniel~Wigger}
\email{daniel.wigger@pwr.edu.pl}
\affiliation{Department of Theoretical Physics, Wroc\l{}aw University of Science and Technology, 50-370~Wroc\l{}aw, Poland}

\author{Matthias~Wei\ss{}}
\affiliation{Lehrstuhl f\"ur Experimentalphysik 1 and Augsburg Centre for Innovative Technologies (ACIT), Universit\"at Augsburg, 86159 Augsburg, Germany}
\affiliation{Institute of Physics, University of M\"unster, 48149 M\"unster, Germany}

\author{Michelle~Lienhart}
\affiliation{Lehrstuhl f\"ur Experimentalphysik 1 and Augsburg Centre for Innovative Technologies (ACIT), Universit\"at Augsburg, 86159 Augsburg, Germany}

\author{Kai~M\"uller}
\affiliation{Walter Schottky Institut, Technische Universit\"at M\"unchen, 85748~Garching, Germany}

\author{Jonathan~J.~Finley}
\affiliation{Walter Schottky Institut, Technische Universit\"at M\"unchen, 85748~Garching, Germany}

\author{Tilmann~Kuhn}
\affiliation{Institut of Solid State Theory, University of M\"unster, 48149 M\"unster, Germany}

\author{Hubert~J.~Krenner}
\affiliation{Lehrstuhl f\"ur Experimentalphysik 1 and Augsburg Centre for Innovative Technologies (ACIT), Universit\"at Augsburg, 86159 Augsburg, Germany}
\affiliation{Institute of Physics, University of M\"unster, 48149 M\"unster, Germany}

\author{Pawe\l{}~Machnikowski}
\affiliation{Department of Theoretical Physics, Wroc\l{}aw University of Science and Technology, 50-370~Wroc\l{}aw, Poland}

\date{\today}

\begin{abstract}
Quantum technologies that rely on photonic qubits require a precise controllability of their properties. For this purpose hybrid approaches are particularly attractive because they offer a large flexibility to address different aspects of the photonic degrees of freedom. When combining photonics with other quantum platforms like phonons, quantum transducers have to be realized that convert between the mechanical and optical domain. Here, we realize this interface between phonons in the form of surface acoustic waves (SAWs) and single photons, mediated by a single semiconductor quantum dot exciton. In this combined theoretical and experimental study, we show that the different sidebands exhibit characteristic blinking dynamics that can be controlled by detuning the laser from the exciton transition. By developing analytical approximations we gain a better understanding of the involved internal dynamics. Our specific SAW approach allows us to reach the ideal frequency range of around 1~GHz that enables simultaneous temporal and spectral phonon sideband resolution close to the combined fundamental time-bandwidth limit.
\end{abstract}

\maketitle

\section{Introduction}
Quantum emitters are arguably one of the most important elements in photonic quantum technologies~\cite{OBrien2009,Elshaari2020}. A promising approach in quantum applications is based on time-bin encoding~\cite{horodecki2009quan, pan2012mult}, which requires a precise control over the spectral and temporal degrees of freedom of single photons. Naturally, such reconfiguration of quantum emitters is demanding because it requires a high-level of control of their spectral properties \cite{ficek1999qua,lukin2020spec} and advanced experimental techniques. Moreover, because time and frequency are inversely connected by the Fourier transform the time-bandwidth limit is a strict constraint for the accessible spectral and temporal resolutions:
$\delta \omega\cdot \delta t \geq 1/2$~\cite{merzbacher1998quantum,busch1990ener}.
Consequently, studies are either performed in the frequency domain ($\delta
\omega\rightarrow 0$) or in the time domain ($\delta t\rightarrow 0$) using for instance
resonance fluorescence \cite{Flagg2009,Pingault2017} and ultrafast four-wave-mixing
techniques \cite{Fras2016}, respectively. The intermediate regime with resolutions
satisfying $\delta \omega\cdot \delta t \approx 1/2$ is of particular interest, because
here the time evolution of 
individual spectral components can be studied with the highest accuracy allowed by the
fundamental laws of physics. This promises deeper understanding of switching dynamics with a combined access to both, spectral and temporal properties. Only by reaching this ideal control on the single photon level will allow to operate single solid state qubits as fully-fledged optomechanical quantum transducers.

Here, we report on a combined theoretical and experimental study on the real-time spectral
dynamics of a dynamically modulated quantum emitter near the time-bandwidth limit. We develop
a theory of coherent light scattering from a quantum emitter whose transition
energy experiences an arbitrary temporal modulation in the limit of weak optical
excitation. Our general theoretical predictions are compared to an
experiment in which we induce dynamics of spectral components by interfacing the excitonic
two-level system of a single semiconductor quantum dot (QD) as a model quantum
emitter simultaneously with the coherent phonon field of a surface acoustic wave (SAW) and a
coherent laser beam. The SAW induces a periodic modulation of the narrow excitonic
transition which, in the frequency domain, leads to the generation of phononic sidebands
(PSBs) in the scattered photon spectrum \cite{metcalfe2010reso, Golter2016, villa2017surface, weiss2021opto},
showing anti-bunched character of the scattered light \cite{villa2017surface, weiss2021opto}. 
By advancing our resolved PSB spectroscopy on single SAW-modulated QDs to a fully-fledged
time-domain detection scheme and introducing parametric detuned optical excitation we are
able to demonstrate controlled photon scattering with time and frequency resolution very close to the fundamental time-bandwidth limit.

\section{Experiment}
\subsection{Sample design and experimental setup}
\begin{figure*}[t]
	\centering
	\includegraphics[width=1.85\columnwidth]{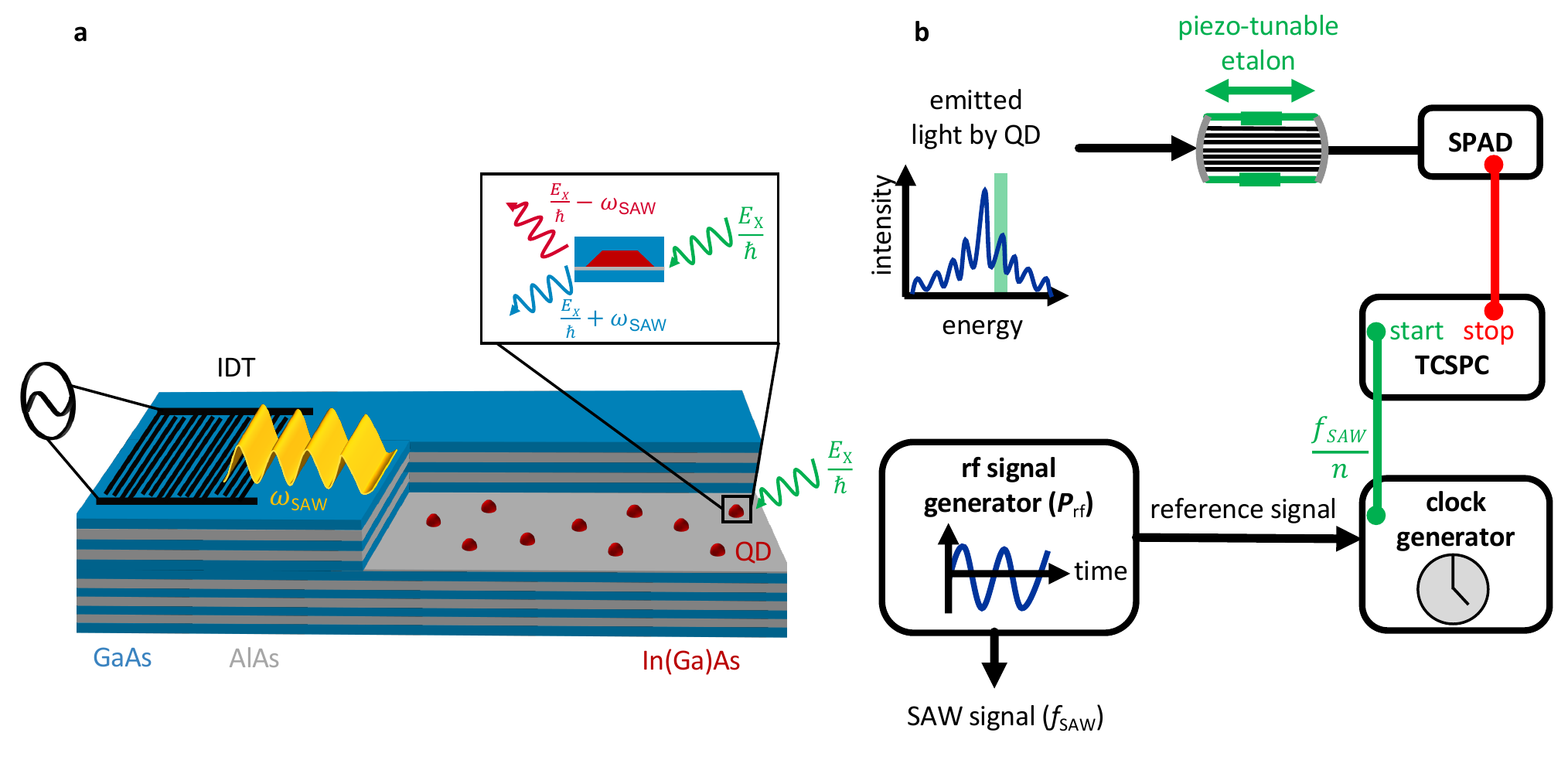}
	\caption{(a) Schematic of the used sample. It includes a layer of self-assembled QDs embedded in a Bragg-type semiconductor microcavity. The QD exciton transition is driven by a resonant laser which is tuned in resonance with its exciton transition energy $E_{\rm X}$. Simultaneously, SAWs are generated by an IDT on the surface of the sample to dynamically strain the positioned single QD.
		(b)  Experimental setup to measure the formation of PSBs in the time domain. The transmission energy of the etalon is changed in discrete steps and for each step a time-transient is recorded using TCSPC.}
	\label{fig:expsetup}
\end{figure*}%
The investigated sample was grown by molecular beam epitaxy and consists of a GaAs/AlAs Bragg-type microcavity with a layer of self-assembled In(Ga)As QDs in between as schematically depicted in Fig.~\ref{fig:expsetup}(a). Altogether, the Bragg reflectors contain 8 (or 10) and 15 alternating layers of GaAs and AlAs above and below the single QD-layer, respectively.  The cavity quality factor of $Q \approx 150$ is sufficiently high to increase the light matter interaction but still exclude any time-modulation of the Purcell-effect~\cite{weiss2016surf}. The resonance of the cavity was designed to match the emission band of the QDs and their density is low enough such that a single QD can be addressed optically.  Additionally, the cavity linewidth of $\Delta f \geq 2.2$~THz is sufficiently large to avoid dynamic modulation of the cavity resonance for amplitudes larger than $\Delta f$.

Interdigital transducers (IDTs) with a Split-5-2 configuration were fabricated on the sample surface with electron beam lithography in a lift-off process to enable frequency-tunable excitation of SAWs that are interfaced with the QDs~\cite{schulein2015four,weiss2018mult}. Taking advantage of the wide tunability achievable with our chirped and non-chirped transducer setups ~\cite{weiss2021opto,weiss2018mult} we choose to use SAWs in the frequency band around  $\omega_{\rm SAW}/2\pi\approx 700$~MHz. This is achieved by applying a radio frequency (rf) voltage with a power $P_{\rm rf}$ and frequency $\omega_{\rm rf}$ which is mapped onto the frequency $\omega_{\rm SAW}$ of the resulting SAW.

All our experiments were performed at $T=5$~K employing SAWSs with $f_{\rm SAW}=677.5$~MHz and $P_{\rm rf} = -5$~dBm (Fig.~\ref{fig:dynamics}(a), non-chirped transducer) or $f_{\rm SAW} = 750$~MHz and $P_{\rm rf} = 2$~dBm (Fig.~\ref{fig:detuning}(a) and Fig. \ref{fig:detuning_all_2}(a), chirped transducer), respectively. The values for $P_{\rm rf}$ were chosen to achieve comparable relative SAW amplitudes. Therefore, one has to compensate for both the different IDT efficiencies at different designs and frequencies, and the differing strength of acousto-optical coupling at different frequencies by the choice of $P_{\rm rf}$.

The acoustic field generated by the IDTs propagates along the surface over a distance of approximately $1.5$~mm -- $2.0$~mm to the investigated QD which is dynamically strained. The oscillating deformation potential induces an $\omega_{\rm SAW}$-periodic spectral modulation of the optical transition of the amplitude $\hbar \Delta$ which is given by $E_{\rm X}+\hbar \Delta \cdot \cos(\omega_{\rm SAW}t)$. In our experiment any additional modulation by the SAW due to a piezoelectrically induced Stark-shift can be excluded~\cite{weiss2014dyna}.

The QD exciton transition is excited by a narrow band continuous wave laser ($\omega_{\rm laser}/2\pi\approx \omega_{\rm exciton}/2\pi\approx 330$~THz, linewidth $\delta \omega_{\rm laser}/2\pi \leq 100$~kHz) driven in the limit of low Rabi frequencies reaching the regime of single photon operation.

Furthermore, we are able to add an optical detuning $\Delta_{\rm opt}=\omega_{\rm laser}-\omega_{\rm exciton}$ between the QD transition $\omega_{\rm exciton}$ and the laser $\omega_{\rm laser}$, which will be used as an external tuning parameter for the PSBs' spectral dynamics. The tuning of the laser frequency is experimentally implemented by piezo tuning the length of the laser cavity.

The laser photons scattered from the QD are collected by the same objective and spectrally filtered by a piezo-tunable Fabry Perot etalon (free spectral range $ \Delta \nu_{\rm FSR}=60$~GHz, finesse $\mathcal{F}=263$, spectral resolution $\delta \nu = \delta\nu_{\rm FSR}/\mathcal{F}=227$~MHz) as shown in Fig.~\ref{fig:expsetup}(b). The resonance fluorescence signal is then detected by a single photon avalanche photodetector (SPAD). Excitation and detection are cross-polarized in order to suppress reflected laser light.

\subsection{Time domain spectroscopy}
In this work we investigate the evolution of PSBs in the time domain, which is experimentally realized as depicted in Fig.~\ref{fig:expsetup}(b). Here, the etalon is scanned in discrete energy steps and time transients are recorded for every step using time-correlated single photon counting (TCSPC). The reference signal for the time correlated single photon counting is provided by a clock generator. It is set to an integer fraction of the SAW frequency $f_{\rm SAW}/n$ which enables that any periodic process in the time domain, whose periodicity is coupled to the SAW period $T_{\rm SAW}$, can be resolved. It is important to note that the time resolution $\Delta t$ of the experiment is restricted by the time-bandwidth limit. Therefore, the maximum time resolution is determined by the spectral resolution of the etalon $\delta \nu = 227$~MHz with $\Delta t \geq (4\pi \delta \nu)^{-1}=350$~ps. In order to best resolve the SAW induced dynamics, the SAW frequency generated by the IDT was chosen sufficiently low to still resolve sideband splitting. This increases $T_{\rm SAW}$ and ensures a high time resolution of the temporal dynamics. This trade-off is considered in more detail in Appendix~\ref{sec:A_dyn}.

It is important to note that the time resolution $\delta t$ of the experiment is limited by the high spectral resolution of the etalon $\delta\nu=\delta\omega/(2\pi) \approx 225$~MHz. The corresponding temporal resolution is therefore given by $\delta t \geq ( 2\delta \omega)^{-1}\approx 350$~ps. By choosing SAW frequencies in the range $f_{\rm SAW}\le 1$~GHz we have periods of $T_{\rm SAW}\ge 1$~ns, which allows us to resolve simultaneously PSBs in the spectrum and their oscillations in time. Owing to their precise tunability in frequency and amplitude, SAWs provide the ideal platform to push the combined spectral and temporal resolution close to the fundamental time-bandwidth limit.

\section{Theory}
In our model we consider the QD exciton as a two-level system with the ground state $\left| 0\right>$ and the single exciton state $\left| 1\right>$. The corresponding Hamiltonian in the frame rotating with the laser frequency is 
\begin{align}
	H_{\rm S} = \hbar\Delta(t)\left|1\right>\!\left<1\right| + \frac12\hbar
	\Omega(\sigma_+ +   \sigma_-)\, , 
\end{align}
\begin{figure}[t]
	\centering
	\includegraphics[width=\columnwidth]{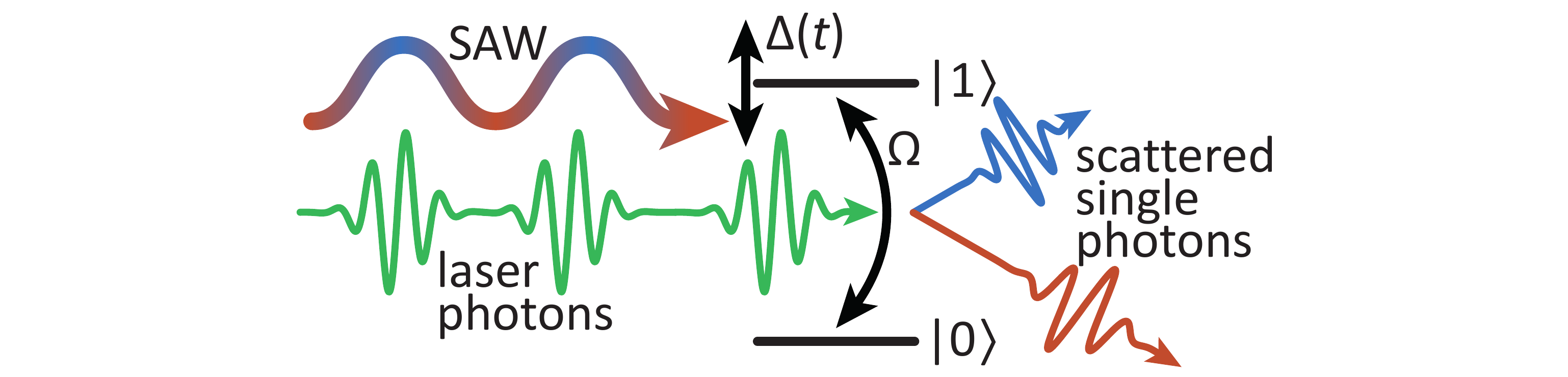}
	\caption{Schematic picture of the considered system. The incoming photons are scattered by the QD exciton while the transition energy is modulated by the SAW. The scattered single photons can gain energy from or loose energy to the SAW field. This results in characteristic phonon sidebands.}
		\label{fig:scheme}
\end{figure}%
where the detuning between exciton transition and laser $\hbar \Delta(t)$ is in general time dependent. In the following we will assume a SAW that modulates the transition energy of the QD exciton via the deformation potential coupling as schematically shown in Fig.~\ref{fig:scheme}~\cite{wigger2017syst}. The Rabi frequency $\Omega$ describes the classical optical driving expressed in terms of the raising and  lowering operators $\sigma_+=\left|1\right>\!\left<0\right|$ and 
$\sigma_-=\left|0\right>\!\left<1\right|$, respectively. The spontaneous decay of the exciton with the rate $\gamma$ is included by the Lindblad dissipator $\mathcal L(\rho_{\rm S}) = \gamma ( \sigma_-\rho_{\rm S}\sigma_+ -  \frac12\{ \sigma_+\sigma_- , \rho_{\rm S} \} )$
and the system's evolution is given by the Master equation in Lindblad form
\begin{align}
	\frac{d}{dt}\rho_{\rm S} = -\frac{i}{\hbar}[H_{\rm S},\rho_{\rm S}] + \mathcal L(\rho_{\rm S})\, .
	\label{eq:eom}
\end{align}
To retrieve the measured resonance fluorescence signal we calculate the two-time correlation function 
\begin{align}\label{eq:G}
	G^{(1)}(t,t+\tau) = \left< \sigma_+(t)\sigma_-(t+\tau) \right>\,.
\end{align}
To account for the resonance fluorescence in the weak excitation limit this function needs to be evaluated in the leading order in the laser field. Assuming $t$ large enough for the system to be in a steady state, the Markovian dynamics can be calculated using the Lax regression theorem \cite{Meystre2007} as recently used in Ref.~\cite{weiss2021opto}. There, it was shown that the correlation function for an arbitrary energy modulation $\hbar\Delta(t)$, in our case induced by a SAW field, and for $t\gg 1/\gamma$ (steady state), reads to the leading order in $\Omega$
\begin{align}\label{eq:G1}
	G^{(1)}(t,t+\tau) & =  \frac{\Omega^{2}}{4}
			\int\limits_{-\infty}^{t} {\rm d}s\,e^{-(\gamma/2)(t-s)+i\Phi(s,t)} \notag\\
			&\times \int\limits_{-\infty}^{t+\tau}{\rm d}s'\,e^{-(\gamma/2)(t+\tau-s')-i\Phi(s',t+\tau)}\, ,
\end{align}
with the time dependent phase function $\Phi(t_0,t) = \int\limits_{t_0}^t \Delta(s) {\rm d}s$.

The action of the spectral filter (Fabry-Perot etalon) can be included leading to the
scattered and filtered intensity dynamics depending on the frequency shift of the
filter~$\omega_s$~\cite{weiss2021opto} 
\begin{align}\label{eq:spectrum}
		&I_{\rm F}(t,\omega_s) \\&=  {\rm Re}\!\! \int\limits_{-\infty}^{\infty} \!\!{\rm d}\tau \!\!\int\limits_0^\infty \!\! {\rm d}s \,
		F^*(t-\tau)F(t-\tau-s)  G^{(1)}(\tau,\tau+s) e^{i\omega_s s} \! ,\notag
\end{align}
where $F(t)$ is the Fourier transform of the spectral filter $\tilde{F}(\omega)$ and is assumed as an exponential decay with rate $1/\delta\omega$, in agreement with the Lorentzian frequency distribution of the etalon used in the experiment. In our numerical simulations that are directly compared to the experiment we simply implement Eqs.~\eqref{eq:G1} and \eqref{eq:spectrum} with a given energy modulation $\hbar\Delta(t)$ entering in $\Phi(t_0,t)$. Here we consider a single harmonic SAW modulation of the transition energy in the form
\begin{subequations}\begin{align}\label{eq:Delta}
	\Delta(t) &= \Delta_0 \cos(\omega_{\rm SAW} t)\,,
\end{align}
hence
\begin{align}
	\Phi(t_0,t) &= D\big[ \sin(\omega_{\rm SAW} t) - \sin(\omega_{\rm SAW} t_0) \big]\, ,
\end{align}\end{subequations}
where $D=\Delta_0/\omega_{\rm SAW}$ is the maximal strain-induced energy shift $\Delta_0$ relative to the SAW frequency $\omega_{\rm SAW}$. As shown in Appendix~\ref{sec:A_spec_time} Eq.~\eqref{eq:spectrum} can be further evaluated leading to
\begin{subequations}
\begin{align}\label{eq:spectrum_dyn}
		I_{\rm F}(t,\omega_s) &=  \frac12 \left| \frac{1}{2\pi} \sum_{n=-\infty}^\infty b_n e^{-in\omega_{\rm SAW} t} \tilde{F}(n\omega_{\rm SAW} - \omega_s) \right|^2
\end{align}
with 
\begin{align}\label{eq:b_n}
		b_n &= \Omega\pi (-i)^n \\
		&\times \int\limits_0^\infty {\rm d}u\ e^{-(\gamma/2)u}J_n\left[2D\sin\left(\frac12 \omega_{\rm SAW}u\right)\right]e^{in\frac12 \omega_{\rm SAW}u}\,,\notag
\end{align}
\end{subequations}
where $J_n$ is the $n$-th Bessel function of 1st kind. The spectral dynamics are periodic in time and we can directly give the time-integrated scattering spectrum
\begin{align}\label{eq:spectrum_int}
		\bar{I}_{\rm F}(\omega_s) &= \frac{1}{4\pi} \sum_{n=-\infty}^\infty \left|  b_n \tilde{F}(n\omega_{\rm SAW} - \omega_s) \right|^2\,,
\end{align}
which describes a series of peaks equally spaced by $\omega_{\rm SAW}$ and weighted by $b_n$ depending on the SAW amplitude $D$ and the exciton decay rate $\gamma$ as discussed later. The contribution with $n=0$ is the zero phonon line (ZPL) while all other terms are phonon sidebands (PSBs) labeled by the number $n$.

As explained in detail in Appendix~\ref{sec:A_approx} the peak amplitudes can be further approximated under the assumption that $\gamma\ll\omega_{\rm SAW}$ (which is not strictly fulfilled here as discussed in more detail in the the Appendix) and the intensity reads
\begin{align}\label{eq:approx_D}
	\bar{I}_{\rm F}(\omega_s) &\approx \pi^3 \sum_{n=-\infty}^\infty \left| \frac{\Omega}{2}J_{0}(D) J_{n}(D)  \tilde{F}(n\omega_{\rm SAW} - \omega_s) \right|^2\,.
\end{align}
This surprisingly handy expression is tested in Appendix~\ref{sec:A_approx}, demonstrating a reasonable agreement with the full numerically evaluated spectrum. Our analytical expressions slightly deviate from the ones given in Ref.~\cite{metcalfe2010reso} where the authors performed a Fourier transform of the microscopic polarization of the QD exciton and not of the two-time correlation function in Eq.~\eqref{eq:G}.

\section{Results and Discussion}
\subsection{Resonant Excitation}

\begin{figure}[b]
	\centering
	\includegraphics[width=\columnwidth]{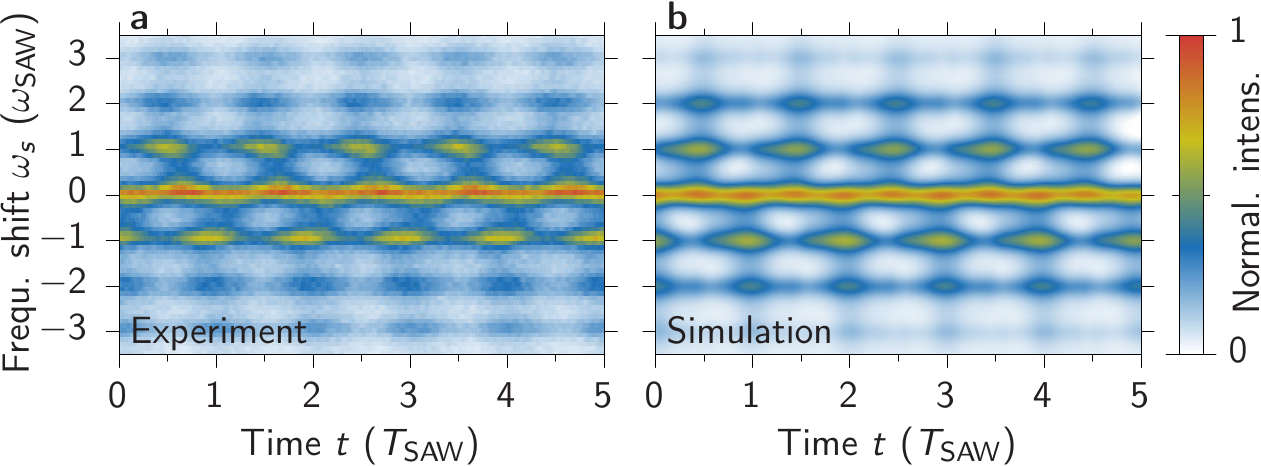}
	\caption{Dynamics of the photon scattering spectrum. (a) experiment, (b) simulation. The experimental parameters are $\omega_{\rm SAW}/2\pi=677.5$~MHz and $P_{\rm rf}=-5$~dBm and for the simulation we use $D=2.7$, $\gamma=1.5\,\omega_{\rm SAW}$, and $\delta\omega=0.325\,\omega_{\rm SAW}$.}
		\label{fig:dynamics}
\end{figure}%
The spectral dynamics of the scattered photons as a function of the frequency passing the etalon $\omega_s$ is depicted in Fig.~\ref{fig:dynamics}(a) (additional measurements with different SAW frequencies are shown in Appendix~\ref{sec:A_dyn}). The corresponding numerical simulation based on  Eq.~\eqref{eq:spectrum_dyn} is shown in Fig.~\ref{fig:dynamics}(b) and it immediately renders a great similarity with the experiment. The feature at $\omega_s=0$ shows the resonantly scattered light defining the ZPL and the neighboring ones the respective PSBs. We find that the PSBs strongly oscillate in time and exhibit a perfect anti-phased behavior when comparing PSBs with opposite sign. This phase shift can also be seen in the analytical equation as shown in Appendix~\ref{sec:A_phase}. This shows that the probability of single photon emission in the sidebands is strongly localized in time and its periodicity can easily be tuned over a wide frequency range by the applied SAW frequency. The fact that the PSBs' dynamics can be nicely resolved and that the PSBs are at the same time well separated in energy crucially depends on the ratio between the SAW frequency and the filter width, which is here chosen to $\delta\omega=0.6\,\omega_{\rm SAW}$ in the simulation. On the one hand a broader filter would increase the temporal resolution but the PSBs would start to overlap. On the other hand a sharper filter would make the lines sharper in energy but the oscillations in time would get smeared out making them harder to resolve. This reflects the trade-off between the time and frequency resolution and shows that we are close to the fundamental time-bandwidth limit.

\begin{figure}[tb]
	\centering
	\includegraphics[width=0.8\columnwidth]{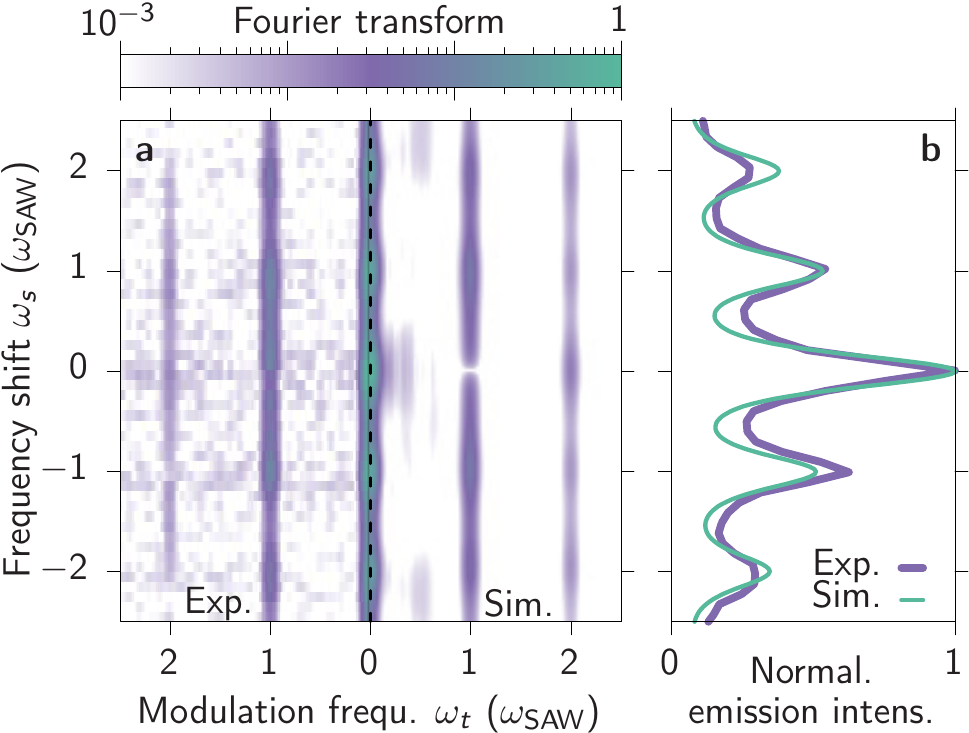}
	\caption{(a) 2D photon scattering spectrum, experiment on the left and simulation on the right hand side. (b) Time-integrated scattering spectra. Model parameters as in Fig.~\ref{fig:dynamics}. A Gaussian filter was used to reach a slight broadening in the $\omega_t$ direction.}
		\label{fig:2D}
\end{figure}

Having a closer look at the dynamical behavior in Fig.~\ref{fig:dynamics}, we realize two
things: (i) that also the ZPL is affected by a slight modulation including the full and
half the SAW period and (ii) that the PSBs' oscillations are not purely harmonic. To
investigate these findings in more detail we Fourier transform the time dependent
scattering spectrum with respect to the real time $t$. The corresponding two-dimensional
(2D) spectra are plotted in Fig.~\ref{fig:2D}(a), where a small artificial broadening in
$\omega_t$ direction was applied. The vertical axis of the 2D spectrum in
Fig.~\ref{fig:2D}(a) remains the energy shift of the scattered light $\omega_s$. The
horizontal axis represents the modulation frequency $\omega_t$ of the spectral
lines. Therefore, it indicates which frequencies contribute to the intensity evolution of
a given spectral line. Note that the spectra as a function of $\omega_t$ are symmetric
with respect to $\omega_t=0$. Therefore we plot the measured spectra on the left hand side
and the calculated ones on the right hand side of $\omega_t=0$.  For completeness in
Fig.~\ref{fig:2D}(b) we show the measured and calculated time-integrated emission
spectra. For the 2D spectra and the time-integrated one experiment and theory agree very
well and the well separated PSBs are symmetric with respect to $\omega_s=0$.

Focusing for
example on the contributions to the ZPL, i.e., a horizontal cut at a frequency shift of
$\omega_s=0$, we find strong components at $\omega_t =0$ and $\omega_t =2\cdot \omega_{\rm
  SAW}$ but a vanishing amplitude at $\omega_t =1\cdot\omega_{\rm SAW}$. This means that
processes including a single phonon cannot contribute to photon emission into the ZPL. The
$\omega_t =2\cdot \omega_{\rm SAW}$ component is in turn increased because processes that
generate and reabsorb a phonon, i.e., forming a loop, result in photon emission into the
ZPL~\cite{weiss2021opto}. 

\subsection{Detuned Excitation}

The integrated emission spectrum shown above is symmetric with respect to the ZPL. This symmetry is broken by introducing a detuning between the laser and the exciton transition $\Delta_{\rm opt}=\omega_{\rm laser}-\omega_{\rm exciton}$. With this, the phase function reads
\begin{align}
	\Phi(t_0,t) = D\big[ \sin(\omega_{\rm SAW} t) - \sin(\omega_{\rm SAW} t_0) \big] + \Delta_{\rm opt}(t-t_0)\, .
\end{align}

\begin{figure}[t]
	\centering
	\includegraphics[width=\columnwidth]{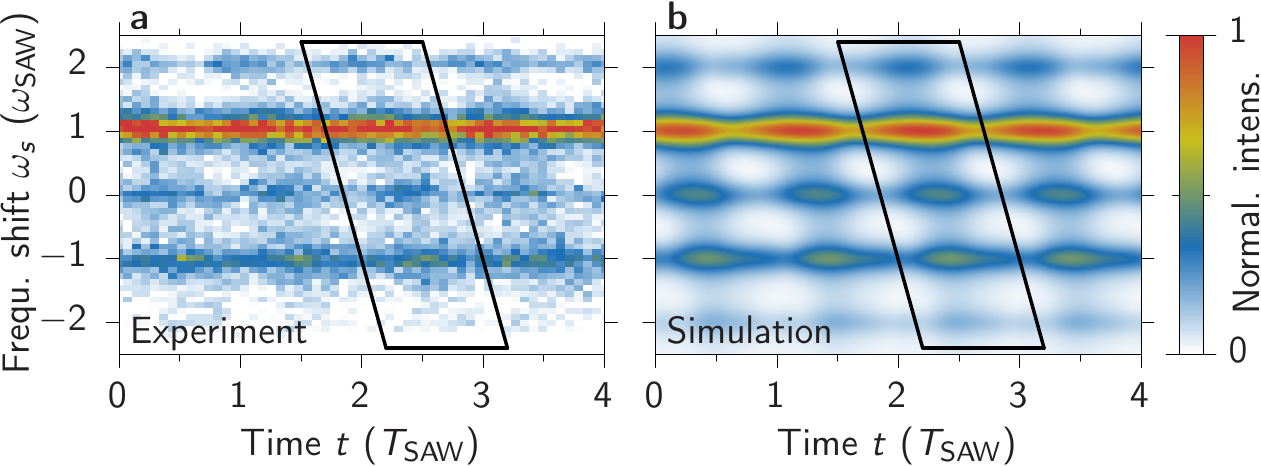}
	\caption{Dynamics of the photon scattering spectrum for a laser detuning of $\Delta_{\rm opt}=\omega_{\rm SAW}$. (a) experiment, (b) simulation. The experimental parameters are $\omega_{\rm SAW}/2\pi=750$~MHz and $P_{\rm rf}=2$~dBm and for the simulation we use $D=1.9$, $\gamma=\omega_{\rm SAW}$, and $\delta\omega=0.3\,\omega_{\rm SAW}$. The red crosses mark maxima at detunings of half SAW frequencies.}
		\label{fig:detuning}
\end{figure}

For Fig.~\ref{fig:detuning} we exemplarily consider a detuning of $\Delta_{\rm opt}=\omega_{\rm SAW}$ and plot the measured time resolved spectrum in (a) and the simulated one in (b). The frequency of the laser agrees with the most intense spectral line at $\omega_s=+1\cdot \omega_{\rm SAW}$, which indicates that still the resonantly scattered light has the strongest contribution in the spectrum. However, the integrated intensity distribution among the PSBs is not symmetric with respect to the laser or the exciton transition ($\omega_s=0$) anymore. While the emission at $\omega_s=0$ is rather week, the PSB at $\omega_s=-1\cdot \omega_{\rm SAW}$ is slightly stronger, although a two-phonon process is required to reach this energy. Obviously, also the beating of the different spectral lines is not anti-phased anymore. Having a close look at the oscillations in Fig.~\ref{fig:detuning} we rather find that a given line reaches a maximum slightly after its neighbor at larger energies and slightly before its neighbor at smaller energies as marked by the parallelograms framing the maxima. This shows that not only the peak intensities but also the relative timing of the oscillations can be controlled by applying a laser detuning with respect to the exciton transition energy. 

\begin{figure}[t]
	\centering
	\includegraphics[width=\columnwidth]{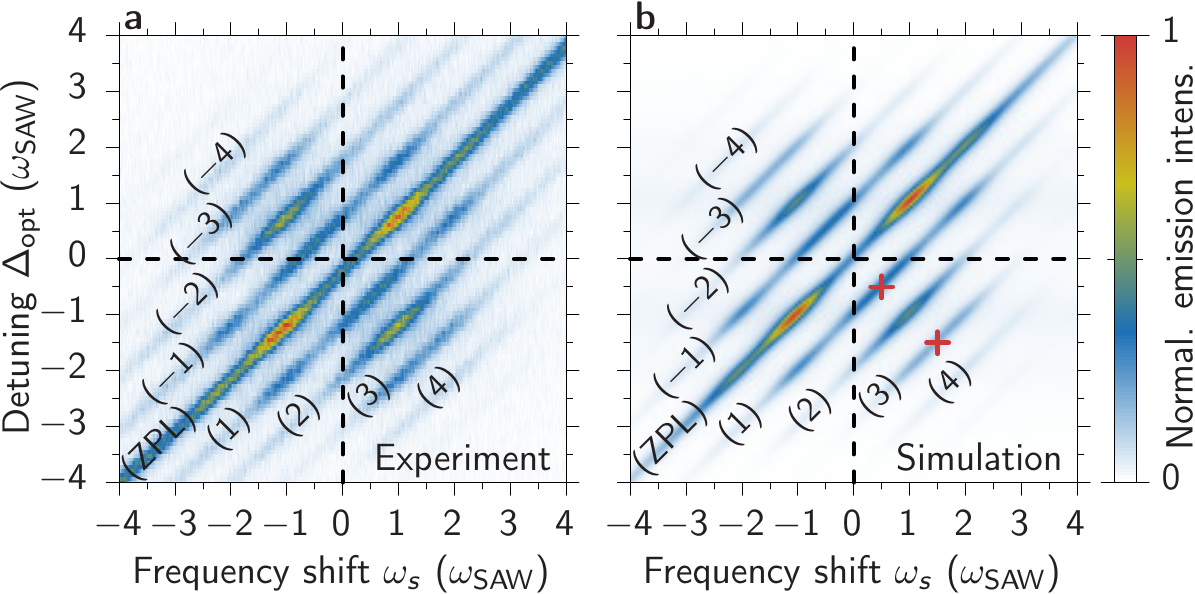}
	\caption{Detuning scan of the time-integrated scattering spectrum. (a) experiment, (b) simulation. The experimental parameters are $\omega_{\rm SAW}/2\pi=1.3$~GHz and $P_{\rm rf}=17$~dBm and for the simulation we use  $D=2$, $\gamma=0.9\,\omega_{\rm SAW}$, and $\delta\omega=0.175\,\omega_{\rm SAW}$.}
		\label{fig:detuning_all}
\end{figure}

To get an overview of this complex interplay between phonon assisted processes and detuning and to explore the full tuning potential of the spectral distribution, we continuously scan the laser detuning $\Delta_{\rm opt}$ and record time-integrated emission spectra. We plot the result in Fig.~\ref{fig:detuning_all}, where the measurement is shown in (a) and the simulation in (b). Note that the spectra in Fig.~5 have been obtained for different parameters compared to Fig.~4, in particular a higher SAW frequency, which leads to more pronounced structures in the spectra. The corresponding spectra for the parameters of Fig.~4 can be found in Appendix~\ref{sec:A_spec}. Again, an excellent agreement of the entire involved pattern is found. The diagonal lines represent the ZPL and the PSBs as indicated by the labels. The overall strongest signal is found when the laser scatters resonantly from the first PSB at $(\omega_s,\Delta_{\rm opt})=(\pm1,\pm1)$. Other local maxima appear at different distinct positions of the plots, mainly when the laser is in resonance with other PSBs, e.g., at $(\omega_s,\Delta_{\rm opt})=(\pm1,\mp 1)$, $(\pm2,\pm 2)$, $(\pm1,\pm 2)$, or $(\pm2,\pm 1)$. All these features are almost exactly the same in experiment and theory. It is also interesting to note that even detunings larger than the maximally reached energy shift, i.e., $\Delta_{\rm opt}>D\,\omega_{\rm SAW}=2\,\omega_{\rm SAW}$, lead to the generation of PSBs. Although of course, the most pronounced peaks are restricted to the range $|\Delta_{\rm opt}|\le 2$.

Having again a look at the peak amplitudes of the scattering spectrum (derived in the same way as for the resonant case)
\begin{align}\label{eq:b_n_det}
		b_n^{\rm (det.)} &= \Omega\pi (-i)^n \int\limits_0^\infty {\rm d}u\ e^{-(\gamma/2)u}\\
		&\times J_n\left[2D\sin\left(\frac12 \omega_{\rm SAW}u\right)\right]e^{i(n\frac12 \omega_{\rm SAW}-\Delta_{\rm opt})u}\,,\notag
\end{align}
we find a peculiar situation for detunings $\Delta_{\rm opt}=\frac12(2m+1)\,\omega_{\rm SAW}$, which are not in resonance with PSBs. Not only for PSB-resonant detunings but also for those placed exactly half way between neighboring PSBs, the complex exponential function in the integrand in Eq.~\eqref{eq:b_n_det} becomes unity for the sideband $n=2m+1$. Because this situation leads to spectral maxima for PSB-resonant excitations this should also hold for half-PSB detunings. This is exactly what we find in Fig.~\ref{fig:detuning_all}(b) at the positions $(\omega_s,\Delta_{\rm opt})=\frac12(2n+1)(\pm1, \mp 1)$ indicating local maxima (examples are marked by red crosses).

The origin of the different efficient or inefficient scattering channels is, that the exciton's coherence dynamics induced by the SAW, leading to the sidebands, is increased or decreased by the detuning, which itself leads to an additional frequency component of the exciton coherence~\cite{hohenester2020}. This shows that the interplay between the dynamics of the phonon-assisted processes and the additional harmonic component from the laser detuning has a non-trivial impact on the photon scattering spectrum. It can be used to access photon channels with half SAW frequency steps opening additional opportunities for frequency and dynamical tuning of single photon emission.

\subsection{Summary and Conclusion}
In summary, our theoretical analysis and experimental demonstration consistently show that
the SAW-induced energy modulations of a single QD exciton can not only 
be utilized to imprint well defined-phonon sidebands on the scattering spectrum of single
photons, but also to induce temporal dynamics of the individual spectral channels of
photon scattering. In particular, the evolution of spectral lines can be controlled by
laser detuning from resonance
and resolved simultaneously in the time and frequency domains down to the fundamental time-bandwidth limit. 
All experimental data are consistent with predictions following from our model. This
proves that our model, which only relies on a time-modulated optically active two-level
system, contains all key features and thus can be applied to other types of quantum
emitters and/or fast tuning mechanisms.

Our findings demonstrate that the application of well-tailored SAW fields and optical
excitations of a single QD exciton can be utilized for a precise temporal and spectral
control of single photon emission. Our approach is an important step in the development of a fully-fledged acousto-optical quantum transducer that allows access to the spectral and temporal domain. This opens new possibilities in the perspective of
hybrid quantum technologies~\cite{xiang2013hybri, kurizki2015quan}, which combine
complementary strengths of dissimilar single systems, while at the same time avoiding
their individual shortcomings. On the one hand semiconductor quantum optics utilizing
single semiconductor QDs has developed a solid understanding and has become
a powerful technological platform~\cite{senellart2017high, hepp2019semi}. On the other
hand phononics, in particular acoustics, has been used in a number of proof-of-principle
studies to manipulate the optical properties of nanosystems~\cite{daly2004imag,
  Bruggemann2011lase, czerniuk2017pico, delsing2019the, wigger2021remo}. Thus, combining
semiconductor quantum optics and phononics, like it is achieved in the present work,
brings a unique hybrid optomechanic quantum chip technology within reach~\cite{ruskov2013chip, pustiowski2015independent, lemonde2018phonon, vogele2020quantum, nysten2020hybrid}. A potential next
step to further improve the controlled timing and energetic channeling of the single
photons would be to use pulsed optical excitations that could be synchronized with the
sideband modulation~\cite{weiss2016surf}.

\section*{acknowledgement}
This project has received funding from Deutsche Forschungsgemeinschaft (DFG, German Research Foundation) via the Emmy Noether Program (KR3790/2) and the Cluster of Excellence “Nanosystems Initiative Munich” (NIM). D.W. thanks the Polish National Agency for Academic Exchange (NAWA) for financial support within the ULAM program (No. PPN/ULM/2019/1/00064). K.M and J.J.F. acknowledge support by Deutsche Forschungsgemeinschaft (DFG, German Research Foundation) under Germany’s Excellence Strategy -- EXC-2111 -- 390814868. K. M. thanks the German Federal Ministry of Education and Research (BMBF) via the funding program Photonics Research Germany (contract number 13N14846) and the Bavarian Academy of Sciences and Humanities (BAdW) for financial support. D.W., K.M., J.J.F., T.K. and P.M. acknowledge support from NAWA under an APM grant. P.M. acknowledges support from the Polish NCN Grant No. 2016/23/G/ST3/04324. M.W., M.L. and H.J.K. thank Achim Wixforth for his continuous support and invaluable discussions.

\appendix

\section{Theoretical details}
\subsection{Derivation of the time resolved spectrum}\label{sec:A_spec_time}
We start from Eqs.~\eqref{eq:G1}, \eqref{eq:spectrum}, and \eqref{eq:Delta}:
\begin{align}\label{eq:G1_SI}
	G^{(1)}(t,t+\tau) & =  \frac{\Omega^{2}}{4}
			\int\limits_{-\infty}^{t} {\rm d}s \,e^{-(\gamma/2)(t-s)+i\Phi(s,t)}\notag\\
			&\times\int\limits_{-\infty}^{t+\tau}{\rm d}s'\,e^{-(\gamma/2)(t+\tau-s')-i\Phi(s',t+\tau)}\, ,
\end{align}
~\\[-1cm]
\begin{align}\label{eq:spectrum_SI}
		&I_{\rm F}(t,\omega) \\
		&\!=\!  {\rm Re}\!\!\left[ \int\limits_{-\infty}^{\infty}\!\! {\rm d}\tau \!\!\int\limits_0^\infty\!\! {\rm d}s\,
		F^*(t-\tau)F(t-\tau-s)  G^{(1)}(\tau,\tau+s) e^{i\omega s} \right]\, , \notag
\end{align}
~\\[-1cm]
\begin{align}
\Phi(t',t) &= D[ \sin(\omega_0 t) - \sin(\omega_0 t') ]
\end{align}
As considered here and in Ref.~\cite{weiss2021opto} $G^{(1)}(t,t+\tau)$ is periodic in $t$ and $\tau$ with the frequency $\omega_0$ (in our case $\omega_{\rm SAW}$) and can therefore be written as
\begin{subequations}
\begin{align}
		G^{(1)}(\tau,\tau+s) & = \frac{1}{(2\pi)^2} \sum_{m,n=-\infty}^\infty c_{mn} e^{-in\omega_0 (\tau+s) }e^{ im\omega_0\tau}\,,
\end{align}
where
\begin{align}
		c_{mn} = \omega_0^2\!\! \int\limits_0^{2\pi/\omega_0}\!\! {\rm d}\tau \!\!\int\limits_0^{2\pi/\omega_0}\!\! {\rm d}\tau'\,
		G^{(1)}(\tau,\tau+\tau')e^{i(n-m)\omega_0\tau}e^{in\omega_0\tau'}\,.
\end{align}
\end{subequations}
Inserting $G^{(1)}$ and introducing the shifted times $u=t-s$ and $u'=t+\tau-s'$ we get
\begin{align}
		c_{mn} &= \omega_0^2 \int\limits_0^{2\pi/\omega_0} {\rm d}\tau \int\limits_0^{2\pi/\omega_0} {\rm d}\tau'\,
		e^{i(n-m)\omega_0\tau}e^{in\omega_0\tau'} \notag\\
		&\times \frac{\Omega^{2}}{4}
			\int\limits_{0}^{\infty} {\rm d}u \,e^{-(\gamma/2)u+i\Phi(\tau-u,\tau)} \notag\\
		&\qquad \times \int\limits_{0}^{\infty} {\rm d}u' \,e^{-(\gamma/2)u'-i\Phi(\tau+\tau'-u',\tau+\tau')} \,.
\end{align}
Using the periodicity of $\Phi$ we can shift the arguments via $\tau'\to\tau'+\tau$ which leads to
\begin{align}
		c_{mn} &= \underbrace{\frac{\Omega\omega_0}{2} \int\limits_0^{2\pi/\omega_0} {\rm d}\tau\,
		e^{-im\omega_0\tau} \int\limits_{0}^{\infty} {\rm d}u \,e^{-(\gamma/2)u+i\Phi(\tau-u,\tau)}}_{=b_m^*} \notag\\
				&\times\underbrace{\frac{\Omega\omega_0}{2} \int\limits_0^{2\pi/\omega_0} {\rm d}\tau'\,
		e^{in\omega_0\tau'} \int\limits_{0}^{\infty} {\rm d}u'\, e^{-(\gamma/2)u'-i\Phi(\tau'-u',\tau')}}_{=b_n^{ }} \notag\\
		&= c_{nm}^* = b_n^{ }b_m^*
\end{align}
We now write Eq.~\eqref{eq:spectrum}/\eqref{eq:spectrum_SI} in the form
\begin{align}
		I_{\rm F}(t,\omega) &= {\rm Re}(z) = \frac12 (z+z^*)\notag\\
		&= \frac12\frac{1}{(2\pi)^2} \sum_{n,m}c_{mn} \int\limits_{-\infty}^\infty {\rm d}\tau\, F^*(t-\tau)\notag\\
		&\qquad \times\int\limits_{0}^\infty {\rm d}s \,F(t-\tau-s)  e^{i\omega s} e^{-in\omega_0(\tau+s)}e^{im\omega_0\tau} \notag \\
		&+  \frac12\frac{1}{(2\pi)^2} \sum_{n,m}c^*_{mn} \int\limits_{-\infty}^\infty {\rm d}\tau\, F(t-\tau)\notag\\
		&\qquad \times\int\limits_{0}^\infty {\rm d}s\, F^*(t-\tau-s)  e^{-i\omega s} e^{in\omega_0(\tau+s)}e^{-im\omega_0\tau}\,,\end{align}
which can again be combined to
\begin{align}
		I_{\rm F}(t,\omega) &= \frac12\frac{1}{(2\pi)^2} \sum_{n,m}c_{mn} \int\limits_{-\infty}^\infty {\rm d}\tau\, F^*(t-\tau)\notag\\
			&\quad \times\int\limits_{-\infty}^\infty {\rm d}s\, F(t-\tau-s)  e^{-i(n\omega_0 -\omega)(s+\tau)} e^{i(m\omega_0-\omega)\tau} \notag\\
						&= \frac12 \left| \frac{1}{2\pi} \sum_{n=-\infty}^\infty b_n e^{-in\omega_0 t} \tilde{F}(n\omega_0 - \omega) \right|^2\,.
\end{align}
This is Eq.~\eqref{eq:spectrum_dyn}.

To calculate the $b_n$ for the harmonic energy modulation from Eq.~\eqref{eq:Delta} in the main text we apply the Jacobi-Anger expansion
\begin{align}
	e^{iz\sin(\theta)} &= \sum_{n=-\infty}^\infty J_n(z)e^{in\theta}\,,
\end{align}
where $J_n$ is the $n$-th Bessel function of 1st kind. This leads to
\begin{align}\label{eq:b_n_SI}
		b_n &= \Omega\pi (-i)^n \int\limits_0^\infty {\rm d}u\ e^{-(\gamma/2)u}\notag\\
		&\quad \times J_n\left[2D\sin\left(\frac12 \omega_{\rm SAW}u\right)\right]e^{in\frac12 \omega_{\rm SAW}u}\\
		&= (-1)^nb_{-n}^* \notag
\end{align}

\subsection{Derivation of the phonon sideband phase shift}\label{sec:A_phase}
As found in the measurement and the numerical simulation in Fig.~\ref{fig:dynamics} the intensities of the phonon sidebands modulate with the SAW frequency. When comparing phonon emission with phonon absorption sidebands we find a phase shift of half a period. This can directly be shown when considering the time-resolved spectrum at inverted frequency
\begin{align*}
	I_{\rm F}(t,-\omega) &= \frac12 \left| \frac{1}{2\pi} \sum_{n=-\infty}^\infty b_n e^{-in\omega_0 t} \tilde{F}(n\omega_0 + \omega) \right|^2\\
		&= \frac12 \left| \frac{1}{2\pi} \sum_{n=-\infty}^\infty (-1)^n b^*_n e^{in\omega_0 t} {\tilde{F}}^*(n\omega_0 - \omega) \right|^2 \\
		&= \frac12 \left| \frac{1}{2\pi} \sum_{n=-\infty}^\infty  b_n e^{-in(\omega_0 t+\pi)} \tilde{F}(n\omega_0 - \omega) \right|^2 \\
		&= I_{\rm F}(t+T_0/2,\omega)\,,
\end{align*}
where we have used that $F(t)$ is real, i.e., $\tilde{F}(-\omega)={\tilde{F}}^*(\omega)$. This relation also directly reveals the symmetry of the time-integrated spectrum with respect to the ZPL.

\subsection{Approximation for resonant excitation}\label{sec:A_approx}
Although we know from our previous studies~\cite{weiss2021opto} that the decay rate is in the same range as the SAW frequency $\gamma\approx \omega_{\rm SAW}$, for the following approximation we ignore the exponential decay in the integrand in Eq.~\eqref{eq:b_n_SI}. This should still be a good approximation when the oscillation of the Bessel function and that of the exponential are much faster than the decay constant $\gamma/\omega_{\rm SAW}$. Then we can perform the integration over one period and use~\cite{gradshteyn}
\begin{align*}
	 \int\limits_{0}^{\pi} J_n\left[ 2D \sin (x)\right] e^{in x} \,{\rm d}x &= \pi i^nJ_{0}(D)J_{n}(D)\, .
\end{align*}
With this we finally find the approximated intensity of the spectral line with an energy shift of $\hbar n\omega_{\rm SAW}$
\begin{align}\label{eq:approx_D_SI}
	b_n &\approx \Omega \pi^2 J_{0}(D)J_{n}(D) \,.
\end{align}
%
\begin{figure}[tb]
	\centering
	\includegraphics[width=0.8\columnwidth]{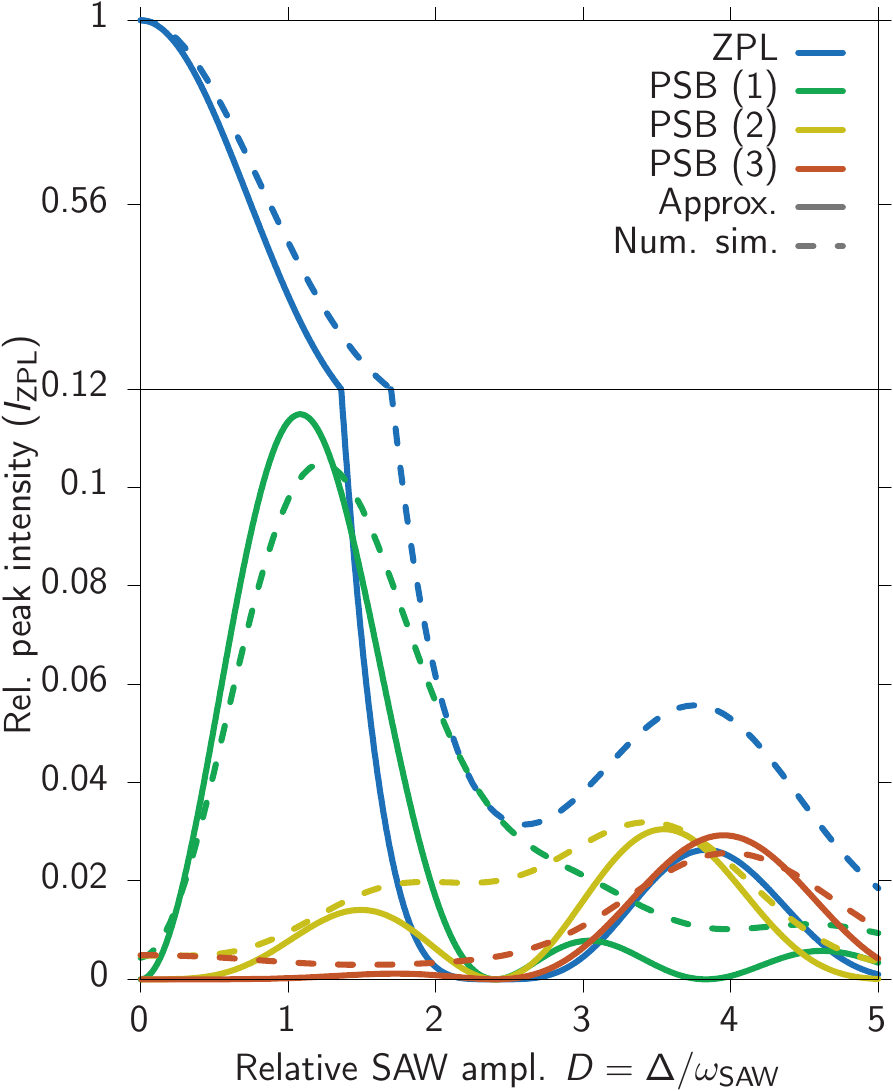}
	\caption{Considered system parameters: $\gamma=\omega_{\rm SAW}$, $\delta\omega=0.1\,\omega_{\rm SAW}$.}
		\label{fig:approx_D}
\end{figure}
This approximation is tested in Fig.~\ref{fig:approx_D} where we compare Eq.~\eqref{eq:approx_D_SI} (solid lines) with numerically calculated spectra (dashed lines) that also include a spectral filter of Lorentzian width $0.1\omega_\text{SAW}$. This filter leads to a broadening and a partial overlap of the different spectral lines in the numerical evaluation. This additionally weakens the strong approximation performed in Eq.~\eqref{eq:approx_D_SI}. Still the position and the height of the maxima in the approximation agree surprisingly well with the numerical calculation. Therefore we can conclude that the analytic evaluation of the $G^{(1)}$ function in Eq.~\eqref{eq:G1_SI} without additional filtering is a reasonable way to evaluate the SAW-induced spectral dynamics. However, to perform actual comparisons with the experiments we perform the simulations numerically. A closer look reveals that the deviations between the solid lines and the dashed lines of the respective color are smaller for larger sideband numbers. While the blue curves (ZPL, $n=0$) agree very well for small $D$, they have a larger deviation at the second maximum around $D=4$. At the same time the orange curves [PSB (3), $n=3$] are almost the same around $D=4$. Overall we find that the approximation works reasonably well around the main maxima of the curves, while around small values of the full model larger relative deviations occur.

\section{Auxiliary experiments}
\subsection{Additional time-resolved measurements}\label{sec:A_dyn}
\begin{figure}[tb]
	\centering
	\includegraphics[width=\columnwidth]{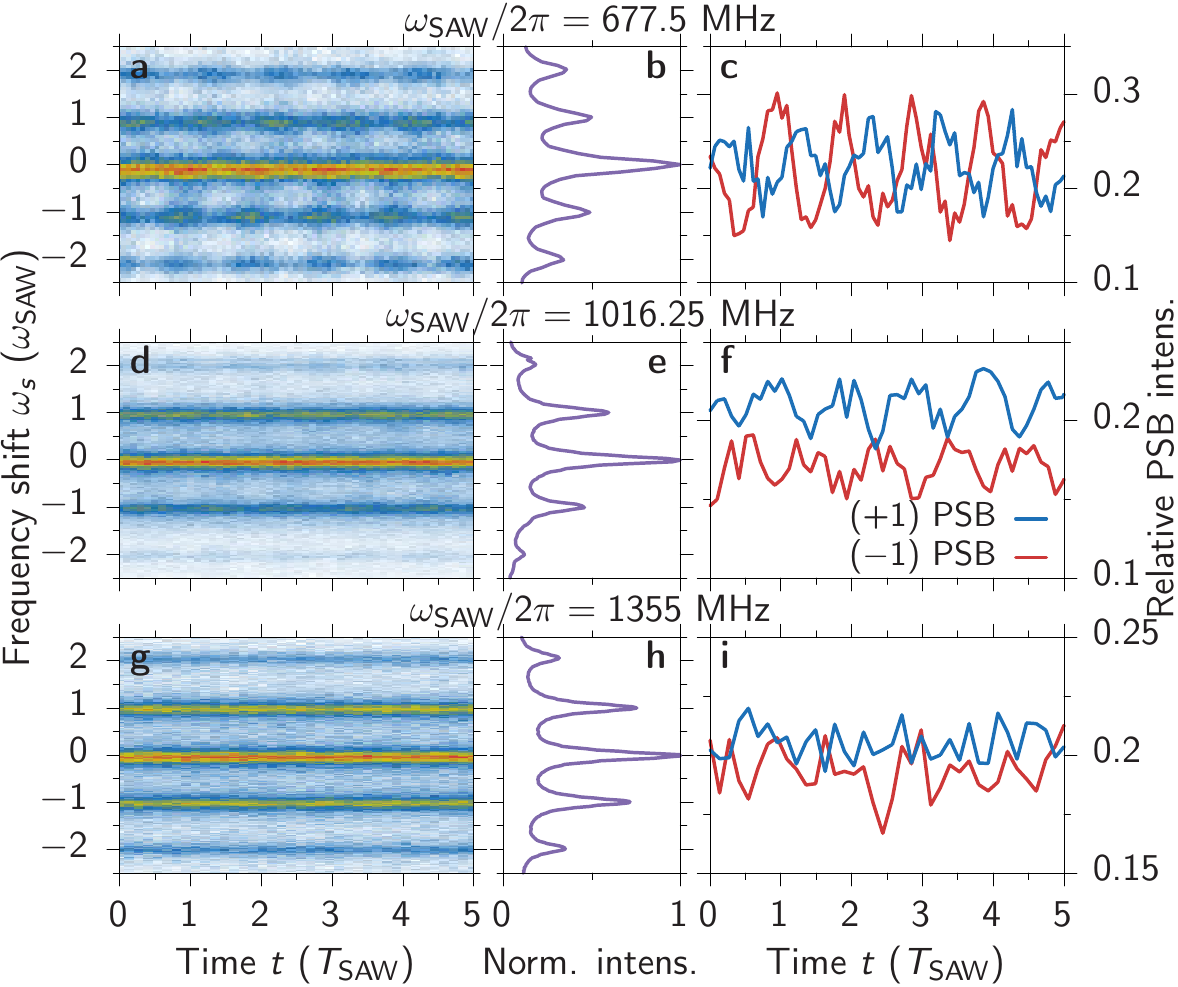}
	\caption{(a, d, g) Time and energy resolved resonance fluorescence intensity for a dynamically strained QD. (b, e, h) Time-integrated resonance fluorescence intensity, obtained by integrating the data shown in (a, d, g) over time. (c, f, i) Time dependent intensities for the ($\pm 1$) PSBs. Measurements are shown for three different SAW frequencies $\omega_{\rm SAW}/2\pi=677.5$~MHz (a, b, c), $1016.25$~MHz (d, e, f) and 1355~MHz (g, h, i).}
	\label{fig:tradeOff}
\end{figure}
Figure~\ref{fig:tradeOff} shows time and energy resolved RF emission in (a, d, g), time-integrated RF spectra in (b, e, h), and time-dependent intensity oscillations for the ($\pm 1$) sidebands in (c, f, i) for three different SAW frequencies $f_{\rm SAW}=677.5$~MHz (a, b, c),\linebreak $1016.25$~MHz (d, e, f), and $1355$~MHz (g, h, i) with rf-amplitudes $P_{\rm rf}=-10$~dBm, $-6$~dBm, and $-3$~dBm, respectively. For the lowest frequency shown here (a, c), clear oscillations of the sideband intensities can be observed in time, as already discussed in the main text. When increasing $\omega_{\rm SAW}$ to about 1~GHz, intensity oscillations can barely be resolved in the spectral dynamics shown in Fig.~\ref{fig:tradeOff}(d). Only a closer examination of the extracted intensities for the ($\pm 1$) PSBs in (f) reveals a clear, although much weaker, anti-correlated oscillation in time. In contrast, for the highest frequency of $\omega_{\rm SAW}/2\pi = 1.355$~GHz in (g, i), no oscillations in time can be resolved. The fact that intensity oscillations cannot be resolved at higher frequencies can be attributed to two effects. First, the radiative decay time of the QD $1/\gamma$ becomes comparable to the SAW period $T_{\rm SAW}$, thus phonons are both emitted and absorbed within one photon emission process. This limits the fundamental resolution of this measurement technique to the decay time $1/\gamma$ of the QD. Second, the absolute temporal resolution in units of $T_{\rm SAW}$ of the experimental setup itself decreases with increasing $f_{\rm SAW}$ due to the time-bandwidth limit. The spectral resolution in units of $f_{\rm SAW}$ increases with increasing $f_{\rm SAW}$, which can clearly be recognized by the time-integrated spectra presented in Figs.~\ref{fig:tradeOff}(b, e, h). Here, a clear increase of the spectral resolution with increasing frequency $\omega_{\rm SAW}$ and thus PSB splitting can be observed. While, for the two higher frequencies, neighboring PSBs are clearly separated, for the lowest frequency already a clear overlap is recognizable. These measurements illustrate the trade-off that must be made between spectral and temporal resolution. This is caused by the high spectral resolution of the etalon that limits the temporal resolution to about $\delta t = 350\,$ps. Decreasing the SAW frequency $\omega_{\rm SAW}$ makes it possible to increase the relative temporal resolution $\delta t/T_{\rm SAW}$, while the sideband splitting, and thus the relative spectral resolution, is decreased. The exact opposite applies to an increase of the SAW frequency. Therefore the SAW frequency can be adjusted to match the requirements for a specific experiment.

\subsection{Additional detuning scan}\label{sec:A_spec}
\begin{figure}[h]
	\centering
	\includegraphics[width=\columnwidth]{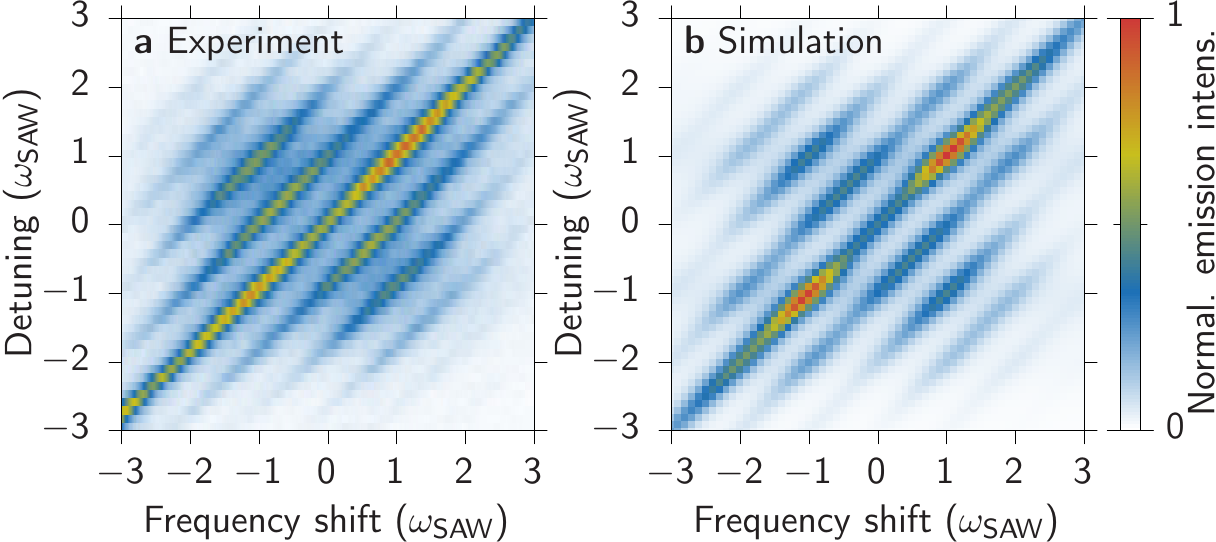}
	\caption{Detuning scan of the time-integrated scattering spectrum. (a) experiment, (b) simulation. The experimental parameters are $\omega_{\rm SAW}/2\pi=750$~MHz and $P_{\rm rf}=2$~dBm and for the simulation we use  $D=1.9$, $\gamma=\omega_{\rm SAW}$, and $\delta\omega=0.3\,\omega_{\rm SAW}$.}
		\label{fig:detuning_all_2}
\end{figure}

In Fig.~\ref{fig:detuning_all_2} we show the full detuning scan from the exemplary spectral dynamics shown in Fig.~\ref{fig:detuning}. The representation is the same as for Fig.~\ref{fig:detuning_all}. The pattern of spectral maxima looks slightly different from the one in the main text because of the different excitation and decay conditions determined by $\omega_{\rm SAW}$, $P_{\rm rf}$, and $\gamma$. However, we still find a good agreement between measurement and simulation confirming the chosen system parameters. It also demonstrates again how the spectral properties of the scattered photons can be fine-tuned by choosing the detuning between laser and QD exciton transition.

%

\end{document}